\documentclass[aps,prl,twocolumn,floatfix,superscriptaddress,longbibliography]{revtex4-1}

\usepackage{amsmath}
\usepackage{amssymb}
\usepackage{times}
\usepackage{braket}
\usepackage[pdftex]{graphicx}
\usepackage{blindtext}
\usepackage{color}

\renewcommand{\vec}[1]{\boldsymbol{#1}}

\newcommand{\change}[1]{\textcolor{black}{#1}}
\newcommand{\changeb}[1]{\textcolor{black}{#1}}
\newcommand{\changec}[1]{\textcolor{black}{#1}}
\newcommand{\changed}[1]{\textcolor{black}{#1}}

\begin{document}

\title{Magnetoelectric effect and orbital magnetization in skyrmion crystals: Detection and characterization of skyrmions}

\author{B{\"o}rge G{\"o}bel}
\email[Corresponding author. ]{bgoebel@mpi-halle.mpg.de}
\affiliation{Max-Planck-Institut f\"ur Mikrostrukturphysik, D-06120 Halle (Saale), Germany}

\author{Alexander Mook}
\affiliation{Max-Planck-Institut f\"ur Mikrostrukturphysik, D-06120 Halle (Saale), Germany}

\author{J\"urgen Henk}
\affiliation{Institut f\"ur Physik, Martin-Luther-Universit\"at Halle-Wittenberg, D-06099 Halle (Saale), Germany}

\author{Ingrid Mertig}
\affiliation{Max-Planck-Institut f\"ur Mikrostrukturphysik, D-06120 Halle (Saale), Germany}
\affiliation{Institut f\"ur Physik, Martin-Luther-Universit\"at Halle-Wittenberg, D-06099 Halle (Saale), Germany}

\date{\today}

\begin{abstract}
Skyrmions are small magnetic quasiparticles, which are uniquely characterized by their topological charge and their helicity. In this Rapid Communication, we show via calculations how both properties can be determined without relying on real-space imaging. The
orbital magnetization and topological Hall conductivity measure the arising magnetization due to the circulation of electrons in the bulk and the occurrence of topologically protected edge channels due to the emergent field of a skyrmion crystal. Both observables quantify the topological Hall effect and distinguish skyrmions from antiskyrmions by sign. 
Additionally, we predict a magnetoelectric effect in skyrmion crystals, which is the generation of a magnetization (polarization) by application of an electric (magnetic) field. This effect is quantified by spin toroidization and magnetoelectric polarizability. The dependence of the transverse magnetoelectric effect on the skyrmion helicity fits that of the classical toroidal moment of the spin texture and allows to differentiate skyrmion helicities: it is largest for Bloch skyrmions and zero for N\'{e}el skyrmions. We predict distinct features of the four observables that can be used to detect and characterize skyrmions in experiments.
\end{abstract}


\maketitle

\begin{figure}
  \centering
  \includegraphics[width=\columnwidth]{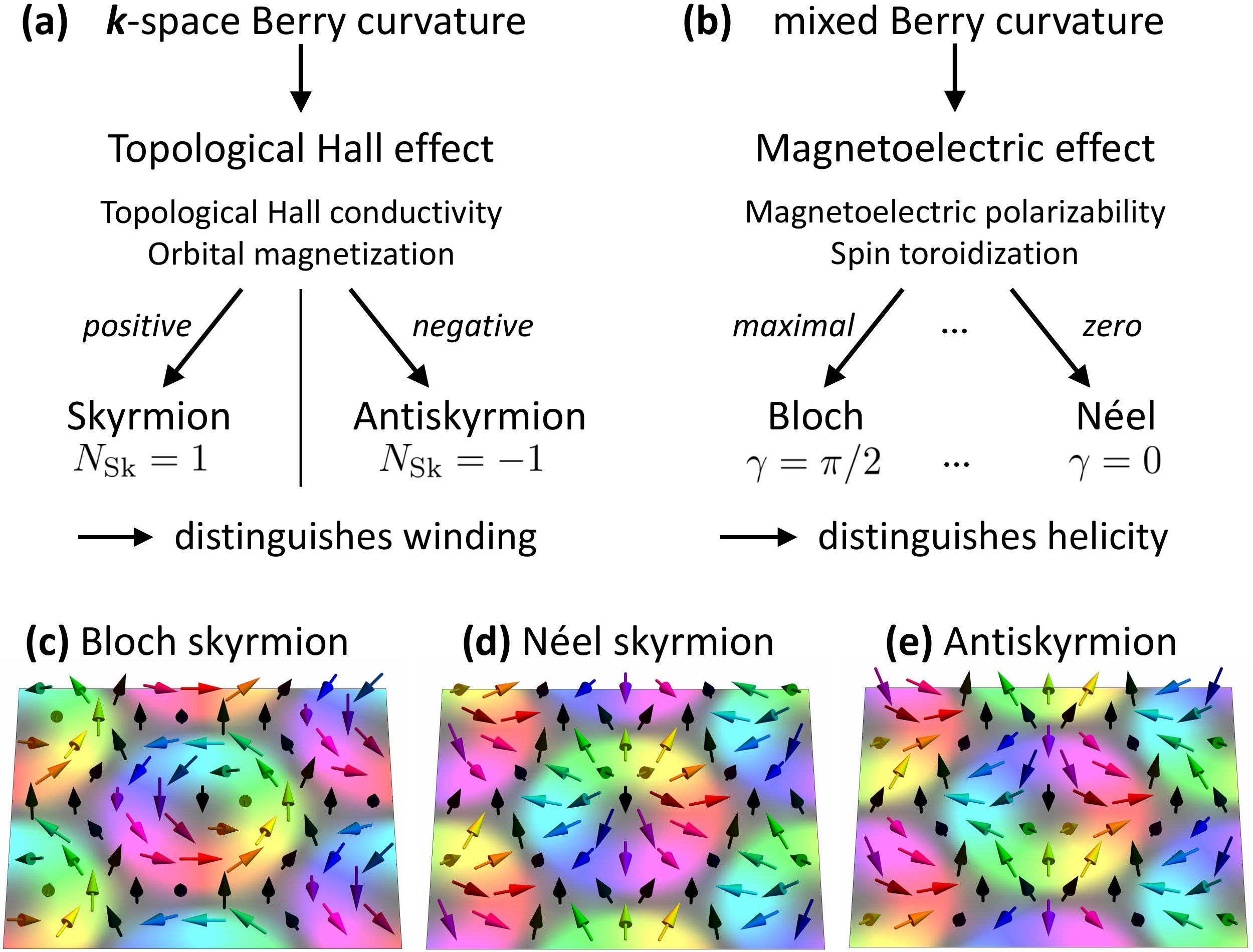}
  \caption{Core message of the paper. (a) Skyrmions and anti-skyrmions are distinguished by the topological Hall effect, (b) the helicity of skyrmions (e.\,g., Bloch and N\'{e}el skyrmions) is differentiated by the magnetoelectric effect. With these quantities Bloch skyrmions (c), N\'{e}el skyrmions (d), and antiskyrmions (e) can be distinguished. The color scale in (c) -- (e) indicates the in-plane orientation of the spins (arrows).}
  \label{fig:skyrmion}
\end{figure}

\paragraph{Introduction.} Skyrmionics has attracted enormous  interest over the recent years, as skyrmions~\cite{skyrme1962unified,bogdanov1989thermodynamically, bogdanov1994thermodynamically,rossler2006spontaneous,muhlbauer2009skyrmion} --- small magnetic quasiparticles, that are topologically protected --- are aspirants to be `bits' in future data storage devices~\cite{fert2013skyrmions,wiesendanger2016nanoscale,romming2013writing,hsu2016electric, zhang2015magnetic,zhang2015magnetic2,jiang2015blowing,boulle2016room, seki2012observation,woo2016observation,gobel2018overcoming}. The integral of the local chirality
\begin{align}
  n_\mathrm{Sk} (\vec{r}) = \vec{s}(\vec{r}) \cdot \left( \frac{\partial \vec{s}(\vec{r})}{\partial x}  \times  \frac{\partial \vec{s}(\vec{r})}{\partial y}  \right)
  \label{eq:spin-chirality}
\end{align}
of a skyrmion with magnetic texture $\vec{s}(\vec{r})$ tells the skyrmion number $N_\mathrm{Sk} = \pm 1$~\cite{nagaosa2013topological,everschor2014real}, that is the topological invariant which characterizes skyrmions and antiskyrmions~\cite{nayak2017magnetic,hoffmann2017antiskyrmions,huang2017stabilization, gobel2018magnetic}, respectively. On top of this, $n_\mathrm{Sk} (\vec{r})$ induces a topological Hall effect (THE)~\cite{neubauer2009topological,schulz2012emergent,kanazawa2011large, lee2009unusual,li2013robust,bruno2004topological,hamamoto2015quantized, gobel2017THEskyrmion,gobel2017QHE,lado2015quantum,ndiaye2017topological, gobel2018family}, which is an additional contribution to the Hall effect~\cite{hall1879new} of electrons in skyrmion crystals (SkXs, a periodic array of skyrmions; Fig.~\ref{fig:skyrmion}).

Another quantity related to the magnetic texture is the orbital magnetization, which is explained in a semi-classical picture by the circulation of conduction electrons in the presence of spin-orbit coupling (SOC)~\cite{chang1996berry,raoux2015orbital,hanke2016role,xiao2005berry,thonhauser2005orbital,ceresoli2006orbital}. Recently, it has been shown that spin chirality, for example in SkXs, can as well induce an orbital magnetization, even without SOC~\cite{dos2016chirality,dias2017insights,lux2017chiral}.

In this \changed{Rapid Communication}, we establish a complete scheme (Fig.~\ref{fig:skyrmion}) for identifying the type of SkX in an experiment, without reverting to real-space imaging (e.\,g., Lorentz microscopy~\cite{yu2010real}).  The TH conductivity and the orbital magnetization describe the THE and are proportional to $N_\mathrm{Sk}$; therefore they differentiate skyrmions from antiskyrmions. Furthermore, we predict a magnetoelectric effect in SkXs, which is within experimental reach; the magnetoelectric polarizability~\cite{essin2009magnetoelectric,essin2010orbital,hayami2014toroidal} and the spin toroidization~\cite{gao2017microscopic,spaldin2008toroidal} allow to determine the skyrmion helicity, by which N\'{e}el skyrmions are differentiated from Bloch skyrmions. While the THE quantities are based on reciprocal space Berry curvature, the magnetoelectric effect is characterized by the mixed Berry curvature analogues (Fig.~\ref{fig:skyrmion}). 

\paragraph{Model and methods.} We consider a two-dimensional \change{square} lattice with a fixed skyrmion texture $\{ \vec{s}_{i} \}$ (unit length, $i$ lattice site). The resulting skyrmions and antiskyrmions can have various helicities [cf. Figs.~\ref{fig:skyrmion}(c)-(e)].

The electrons in the SkX are described by a tight-binding Hamiltonian
\begin{align} 
  H & = \sum_{ij} t \, c_{i}^\dagger \,c_{j} + m \sum_{i} \vec{s}_{i} \cdot (c_{i}^\dagger \vec{\sigma} c_{i}),
  \label{eq:ham_the} 
\end{align}
($c_{i}^\dagger$ and $c_{j}$ creation and annihilation operators, resp.) with Hund's rule coupling. The electron spins interact with the magnetic texture ($m$ coupling energy; $\vec{s}_{i}$ unit vector; $\vec{\sigma}$ vector of Pauli matrices), which could be created by localized $d$ electrons that are not explicitly featured in this one-orbital Hamiltonian.

From the eigenvalues $E_{n}(\vec{k})$ and eigenvectors $\ket{u_{n}(\vec{k})}$ of the Hamiltonian~\eqref{eq:ham_the} we calculate the $\vec{k}$-space and the mixed Berry curvature for band $n$,
\begin{subequations}
\begin{align}
 \Omega_{n}^{(ij)}(\vec{k}) & = -2\mathrm{Im} \braket{\partial_{k_i} u_{n}(\vec{k})|\partial_{k_j} u_{n}(\vec{k})},\\
 D_{n}^{(ij)}(\vec{k}) & = -2\mathrm{Im}\braket{\partial_{k_i} u_{n}(\vec{k}) | \frac{1}{m}\partial_{s_j}u_{n}(\vec{k})},
\end{align}
\end{subequations}
respectively. With
\begin{subequations}
\begin{align}
 v_{nl}^{(j)}(\vec{k}) & \equiv \braket{u_{n}(\vec{k}) | \partial_{k_j} H(\vec{k})| u_{l}(\vec{k})},
 \\
 s_{nl}^{(j)}(\vec{k}) & \equiv \braket{u_{n}(\vec{k}) | \sigma_j| u_{l}(\vec{k})},
\end{align}
\end{subequations}
($j = x, y, z$) we arrive at
\begin{subequations}
\begin{align}
  \Omega_{n}^{(ij)}(\vec{k}) & = -2\mathrm{Im} \sum_{l \ne n}
  \frac{v_{nl}^{(i)}(\vec{k}) v_{ln}^{(j)}(\vec{k})}{[E_{n}(\vec{k}) - E_{l}(\vec{k})]^2},\label{eq:kberry}
  \\
	D_{n}^{(ij)}(\vec{k}) &=  -2\mathrm{Im} \sum_{l \ne n}
  \frac{v_{nl}^{(i)}(\vec{k}) s_{ln}^{(j)}(\vec{k})}{[E_{n}(\vec{k}) - E_{l}(\vec{k})]^2}.\label{eq:mixed}
\end{align}\label{eq:berry}
\end{subequations}
Integration over the occupied states [short-hand notation
$\int_\mathrm{occ} (\cdot) \equiv \sum_{n} \int (\cdot) \, \Theta(E_{n}(\vec{k}) - E_{\mathrm{F}})\,\mathrm{d}^{2}k$
with $E_\mathrm{F}$ Fermi energy and $\Theta$ Fermi distribution at zero temperature] yields the conductivity $\sigma_{ij}$~\cite{nagaosa2010anomalous} and the magnetoelectric polarizability $\alpha_{ij}$~\cite{essin2010orbital,hayami2014toroidal,gao2017microscopic},
\begin{subequations}
\begin{align}
  \sigma_{ij}(E_\mathrm{F}) & = -\frac{e^{2}}{h} \frac{1}{2\pi} \int_\mathrm{occ} \Omega_{n}^{(ij)}(\vec{k}),
  \label{eq:sigma}
  \\
  \alpha_{ij}(E_\mathrm{F}) & = g\mu_{\mathrm{B}}\frac{e}{(2\pi)^2} \int_\mathrm{occ} D_{n}^{(ij)}(\vec{k}).
  \label{eq:alpha}
\end{align}
\end{subequations}
From the orbital magnetic moment~\cite{chang1996berry,raoux2015orbital}
\begin{align}
   \vec{m}_{n}(\vec{k}) & =  -\frac{e}{2\hbar}\,\mathrm{Im} \sum_{l \ne n}
  \frac{\vec{v}_{nl}(\vec{k}) \times \vec{v}_{ln}(\vec{k})}{E_{n}(\vec{k}) - E_{l}(\vec{k})},
\end{align}
we calculate the orbital magnetization~\cite{xiao2005berry}
\begin{align}
\begin{split}
  M_{z}(E_{\mathrm{F}})  = & \frac{1}{(2\pi)^2} \int_\mathrm{occ} m^{(z)}_{n}(\vec{k})+ \frac{e}{\hbar}\frac{1}{(2\pi)^2}
 \\
 \times&\int_\mathrm{occ}\frac{\Omega^{(xy)}_{n}(\vec{k})-\Omega^{(yx)}_{n}(\vec{k})}{2} \, [E_{\mathrm{F}}-E_{n}(\vec{k})];
 \label{eq:orbmag}
\end{split}
\end{align}
likewise, from the spin toroidal moment
\begin{align}
  \vec{t}_{n}(\vec{k}) &= \frac{g\mu_{\mathrm{B}}}{2}\,\mathrm{Im} \sum_{l \ne n}
  \frac{\vec{v}_{nl}(\vec{k}) \times \vec{s}_{ln}}{E_{n}(\vec{k}) - E_{l}(\vec{k})},
\end{align}
the spin toroidization, as recently shown by Gao \textit{et al.}~\cite{gao2017microscopic}
\begin{align}
\begin{split}
T_{z}(E_{\mathrm{F}}) = & \frac{1}{(2\pi)^2} \int_\mathrm{occ} t^{(z)}_{n}(\vec{k}) - g \mu_{\mathrm{B}} \frac{1}{(2\pi)^2}
 \\
 \times &  \int_\mathrm{occ} \frac{D^{(xy)}_{n}(\vec{k})-D^{(yx)}_{n}(\vec{k})}{2} \, [E_{\mathrm{F}}-E_{n}(\vec{k})]. 
 \label{eq:toroid}
\end{split}
\end{align}
The terms with $m^{(z)}_{n}$ and $t^{(z)}_{n}$ capture the intrinsic contributions of each Bloch electron, while the other terms account for the Berry curvatures $\Omega_{n}^{(ij)}$ and $D_{n}^{(ij)}$, which modify the density of states~\cite{xiao2005berry}.

\paragraph{Topological Hall effect as a quantum Hall effect.} Before discussing the novel results concerning  the energy-dependent orbital magnetization, magnetoelectric polarizability, and spin toroidization, a sketch of the band formation and the TH conductivity is adequate; cf.\ Refs.~\onlinecite{gobel2017THEskyrmion} and~\onlinecite{gobel2017QHE}.

For $m = 0$ in the Hamiltonian~\eqref{eq:ham_the}, the so-called zero-field band structure is spin-degenerate because there is neither spin-orbit coupling nor coupling to the SkX magnetic texture.

\begin{figure*}
  \centering
  \includegraphics[width=\textwidth]{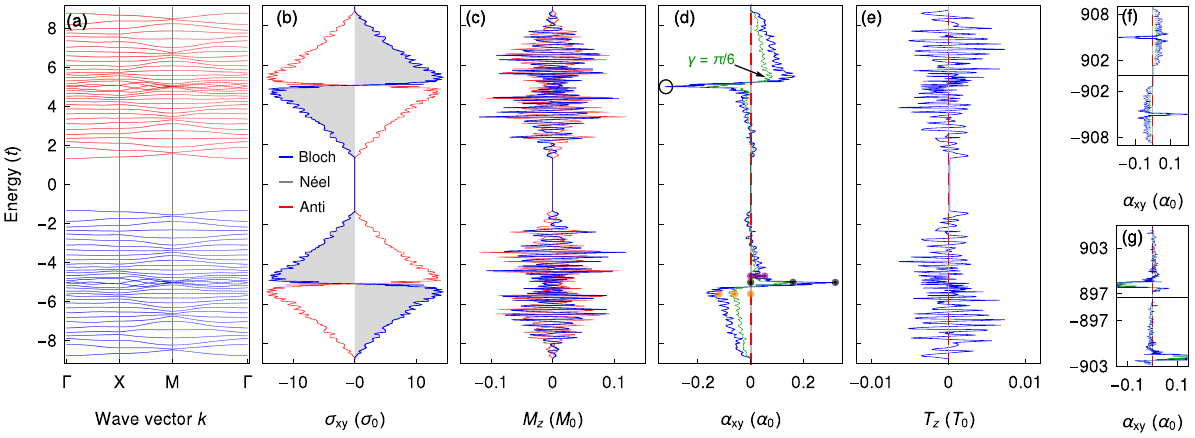}
  \caption{Properties of a skyrmion crystal. Parameters read \change{$n_{\mathrm{b}} = 36$} (sites in the skyrmion unit cell), coupling \change{$m = 5 \, t$}. \change{A Bloch- ($N_{\mathrm{Sk}} = +1$, $\gamma = \pi / 2$; topological charge and helicity), a N\'{e}el- ($N_{\mathrm{Sk}} = +1$, $\gamma = 0$) and an antiskyrmion ($N_{\mathrm{Sk}} = -1$, $\gamma = 0$) are compared}. (a) Band structure, (b) TH conductivity $\sigma_{xy}$, (c) orbital magnetization $M_{z}$, (d) magnetoelectric polarizability $\alpha_{xy}$, and (e) spin toroidization $T_{z}$ are separated into blocks in which the electron spins are aligned parallel (red \change{in a}) or antiparallel (blue) with the skyrmion magnetic texture ($\sigma_0 \equiv e^2 / h$, $M_0 \equiv te / \hbar$, $\alpha_0 \equiv g \mu_{\mathrm{B}} e/at$, and $T_0 \equiv g \mu_{\mathrm{B}}/a$; $a$ is the lattice constant). \change{The band structure is identical for all skyrmion types. In panel (d) results for an intermediate skyrmion with $\gamma=\pi/6$ are shown in addition (green). Colored dots refer to Fig.~\ref{fig:hel}. (f) $\alpha_{xy}$ in the strong-coupling limit $m=900\,t$ and (g) for larger skyrmions $n_b=48$ on a different lattice (triangular).}}
  \label{fig:the}
\end{figure*}

If $m$ is turned on, the spin degeneracy is lifted and the electron spins tend to align locally parallel or antiparallel with the magnetic texture. At $m \approx 5 \, t$ the spin alignment is almost complete and two blocks with $n_{\mathrm{b}}$ (number of sites forming a SkX unit cell) bands each are formed: one for parallel (higher energies) and one for antiparallel alignment (lower energies); \change{see Fig.~\ref{fig:the}a}.

In the limit $m \rightarrow \infty$ the alignment is perfect and the electron spins follow the skyrmion texture adiabatically. Both blocks are identical but shifted in energy. Roughly speaking, besides the rigid shift by $\pm m$, the non-trivial Zeeman term leads to a `condensation' of bands [identified as Landau levels (LLs) in what follows].

The perfect alignment for $m \to \infty$ motivates to transform the Hamiltonian~\eqref{eq:ham_the}: a local spin rotation diagonalizes the Zeeman term~\cite{ohgushi2000spin,hamamoto2015quantized,gobel2017THEskyrmion,gobel2017QHE} and alters the hopping term (the hopping strengths $t_{ij}$ become complex $2\times 2$ matrices). Since the system can be viewed as consisting of two (uncoupled) spin species, it is sufficient to consider only one species. The resulting Hamiltonian \changeb{describes a spin-polarized version of the QHE. Since we discuss charge conductivities the diagonal Zeeman term is dropped and we arrive at the Hamiltonian}
\begin{align}
  H_{\mathrm{\parallel}} & = \sum_{ij} t_{ij}^{(\mathrm{eff})} \, \tilde{c}_{i}^\dagger \,\tilde{c}_{j} 
\end{align}
of a quantum Hall (QH) system (spinless electrons on a lattice)~\cite{hofstadter1976energy,claro1979magnetic, rammal1985landau, claro1981spectrum, thouless1982quantized,hatsugai2006topological, sheng2006quantum}. The effective hopping strengths  $t_{ij}^{(\mathrm{eff})}$ describe the coupling of the electron charges with a collinear inhomogeneous magnetic field
\begin{align}
 B_\mathrm{em}^{(z)}(\vec{r}) & \propto n_\mathrm{Sk}(\vec{r}).
\end{align}
This emergent field~\cite{nagaosa2013topological,everschor2014real} is given by the spin chirality~\eqref{eq:spin-chirality}, that is the real-space Berry curvature in the continuous limit~\cite{everschor2014real,nagaosa2013topological}. The parallel (antiparallel) alignment of the electron spins, corresponding to the upper (lower) block in the band structure for $m \to \infty$, manifests itself in the sign of the nonzero average of $B_\mathrm{em}$.

For finite $m$, the mapping of the THE onto the quantum Hall effect (QHE) --- and the one-to-one identification of bands and LLs --- is reasonable as long as the band blocks are separated, i.\,e., for $m \ge 4 \, t$~\change{\footnote{Please note that in real materials $m \leq 10 \,t$.}}.

The LL character of the bands arises in Chern numbers and in the TH conductivity (Fig.~\ref{fig:the}b). Bands of the upper (lower) block carry Chern numbers of $+1$ ($-1$) due to the positive (negative) average emergent field. As a result, the TH conductivity is quantized in steps of $e^2 / h$. At the shifted van Hove singularity (VHS) \change{$E_\mathrm{VHS} = \pm m$} the TH conductivity changes sign in a narrow energy window.

The quantization and the sign change are closely related to the zero-field band structure~\cite{gobel2017THEskyrmion,gobel2017QHE,gobel2017afmskx}. At the VHS the character of the Fermi lines changes from electron- to hole-like. The bands close to the VHS are simultaneously formed from electron- and hole-like states, leading to a large Chern number that causes this jump.

Having sketched influences of the zero-field band structure on the THE, we derive novel consequences for the orbital magnetization. 

\paragraph{Orbital magnetization.} The block separation manifests itself in the orbital magnetization~\eqref{eq:orbmag} as well. Its energy dependence within the lower block is similar to that in the upper one but with opposite sign (Fig.~\ref{fig:the}c); the latter is explained by the alignment of the electron spin with the magnetic texture.

$M_{z}(E_{\mathrm{F}})$ shows rapid oscillations with zero-crossings within the band gaps,  which is explicated as follows. The emergent field leads to a rotation of an electron wave packet around its center of mass. The first term in Eq.~\eqref{eq:orbmag}, given by $\vec{m}_{n}(\vec{k})$, changes continuously within the bands but is constant within the band gaps. In contrast, the phase-space correction due to the Berry curvature~\cite{xiao2005berry} (second term) varies continuously in energy. Its slope in the band gaps is determined by the TH conductivity,
\begin{align}
 \frac{\partial}{\partial E_\mathrm{F}} M_{z}(E_\mathrm{F}) = \frac{1}{2e} [ \sigma_{yx}(E_{\mathrm{F}}) -\sigma_{xy}(E_{\mathrm{F}})].
 \label{eq:orbmag_slope}
\end{align}
Both gauge-invariant contributions are similar in absolute value but differ in sign. Consequently, their small difference leads to one oscillation per band.

For a better understanding we relate the orbital magnetization in a SkX to that of the associated QH system with (almost) dispersionless bands~\cite{gat2003semiclassical,wang2007orbital,yuan2012orbital}. Besides the oscillations we identify a continuous envelope function~\cite{gat2003semiclassical} (Fig.~S1 in the Supplemental Material~\cite{SupplementalMaterial}). In the SkX this envelope is `deformed' due to the inhomogeneity of the emergent field. Nevertheless, the spectrum of the QH system is quite similar to that of the SkX\@. 

The influences of the two terms in Eq.~\eqref{eq:orbmag} show up `undistorted' in the QH system. The orbital magnetic moment per band (entering the first term) decreases (increases) step-wise at energies below (above) the shifted zero-field VHSs at $E_\mathrm{VHS} \equiv \pm m$ (cf.\ Fig.~S1c). There is no sign change at the VHSs, in contrast to the TH conductivity. Still, a zero-field explanation holds as $\vec{M}$ is also based on the $\vec{k}$-space Berry curvature. At energies below a VHS, LLs are formed from electron-like orbits with a fixed common circular direction. At energies above a VHS, hole-like orbits are formed in addition. Since these exhibit the opposite circular direction, they contribute with opposite sign. Both contributions result in an extremum at the VHS\@.

Size and shape of the orbits dictate the magnitude of the contributions of each band. Therefore, on one hand, the oscillation amplitudes in Fig.~\ref{fig:the}c and Fig.~S1b (corresponding to the step heights in Fig.~S1c) increase with increasing energy distance of Fermi energy and band edges. On the other hand, the oscillation amplitudes vanish at the VHS\@. Recall that the Fermi lines have zero curvature at this particular energy.

When exchanging skyrmions with antiskyrmions the sign of the emergent field changes and so does the sign of both the TH conductivity and the orbital magnetization in Figs.~\ref{fig:the}b and~\ref{fig:the}c, as both characterize the THE\@. These quantities distinguish skyrmions from antiskyrmions but can not distinguish Bloch and N\'{e}el skyrmions. 

\paragraph{Magnetoelectric polarizability.} The independence of all above quantities on the skyrmions' helicity calls for further characterization: this is met by the magnetoelectric effect described by magnetoelectric polarizability and spin toroidization. Both quantities are derived from the mixed Berry curvature $D_{n}^{(ij)}$. \change{If the Fermi energy lies between two Landau levels, the system is insulating. In this case,} the transverse magnetoelectric polarizability
\begin{align}
  \alpha_{xy} & = \left. \frac{\partial M_y}{\partial E_x} \right|_{\vec{B=0}}
  =
 \left. \frac{\partial P_x}{\partial B_y} \right|_{\vec{E=0}} 
 \label{eq:mpeffect}
\end{align}
quantifies the magnetoelectric coupling~\cite{essin2009magnetoelectric} to in-plane fields that are applied to a sample in the SkX phase: an in-plane magnetization $\vec{M}$ (polarization $\vec{P}$) can be modified by an orthogonal in-plane electric field $\vec{E}$ (magnetic field $\vec{B}$~\footnote{The magnetic field $\vec{B}$ is applied in addition to the field that is needed to stabilize the SkX phase.}). If the Fermi energy lies within a Landau level, the system is metallic and cannot exhibit a polarization. Nevertheless, an in-plane magnetization can be produced by perpendicular in-plane currents that are brought about by an applied electric field. This \changed{so called new} magnetoelectric effect in metals \changed{is equivalent to an intrinsic Edelstein effect}~\cite{edelstein1990spin} and  was predicted~\cite{hayami2014toroidal} and confirmed experimentally for UNi$_4$B~\cite{saito2018evidence}, which shows a coplanar toroidal order. \changed{The Onsager reciprocal effect is the inverse Edelstein effect: the generation of a current via the injection of a non-equilibrium spin polarization}.

For a Bloch SkX, the spectrum of the magnetoelectric polarizability $\alpha_{xy}(E_{\mathrm{F}})$, Eq.~\eqref{eq:alpha}, shows sign reversal of the two separated blocks (Fig.~\ref{fig:the}d). Although $\alpha_{xy}$ exhibits \changec{plateaus}, it is not quantized. \change{Around the VHS the curve shows a sharp peak (circle).} 

\change{For $m\gg t$ the spectrum of each block becomes symmetric [Fig.~\ref{fig:the}(f)].} Within a block the sign of $\alpha_{xy}$ mostly remains, in contrast to $\sigma_{xy}$. The monotonicity however is reversed above the VHS, reason being the exchange of $\vec{v}_{ln}$ and $\vec{s}_{ln}$ in Eqs.~\eqref{eq:kberry},~\eqref{eq:mixed}. 
While the sign of the velocity is given by the electron or hole character, the spin is aligned with the magnetic texture, irrespective of the electronic character of band $l$.
The mixing of electron and hole states in a small energy window about $E_\mathrm{VHS}$ leads to a collapse of $\alpha_{xy}$ with a reversed sign for this small energy region. This energy window corresponds to the jump in $\sigma_{xy}$. 
 
\paragraph{Spin toroidization.} Like the magnetoelectric polarizability is related to the TH conductivity, the spin toroidization~\eqref{eq:toroid} is related to the orbital magnetization. It comprises two terms: one given by the spin toroidal moments $\vec{t}_{n}$, the other by the phase-space correction due to the mixed Berry curvature. In analogy to Eq.~\eqref{eq:orbmag_slope}, its slope
\begin{align*}
  \frac{\partial}{\partial E_\mathrm{F}} T_{z}(E_{\mathrm{F}}) & = \frac{1}{2e} [\alpha_{yx}(E_{\mathrm{F}})-\alpha_{xy}(E_{\mathrm{F}})]
\end{align*}
is given by the magnetoelectric polarizability \change{in the band gap}~\cite{gao2017microscopic}.

$T_{z}(E_{\mathrm{F}})$ oscillates rapidly \change{for the Bloch SkX} (Fig.~\ref{fig:the}e). \change{In the strong-coupling limit $m\gg t$ the shape of the oscillations becomes more pronounced}.

\paragraph{Relation to skyrmion helicity.} 

\begin{figure}
  \centering
  \includegraphics[width=1\columnwidth]{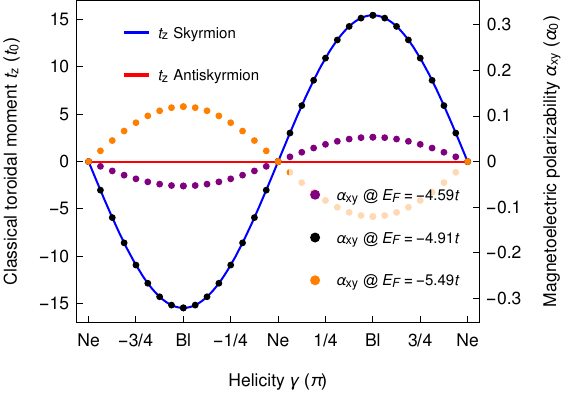}
  \caption{Dependence of the classical toroidal moment $t_{z}$ (blue, red) and the magnetoelectric polarizability $\alpha_{xy}$ on the helicity of a skyrmion for selected Fermi energies $E_{\mathrm{F}}$ (distinguished by color, as indicated; also marked in Fig.~\ref{fig:the}e). $t_{z}$ is proportional to $\alpha_{xy}$, with the proportionality factor depending on $E_{\mathrm{F}}$. $t_0\equiv g\mu_B a$, $\alpha_0 \equiv g \mu_{\mathrm{B}} e/at$.}
  \label{fig:hel}
\end{figure} 

Changing continuously the skyrmion helicity, from Bloch to N\'{e}el skyrmions, $\alpha_{xy}$ and $T_{z}$ are reduced by a Fermi-energy-\emph{independent} factor \change{[Fig.~\ref{fig:hel} and green curve in Fig.~\ref{fig:the}(d)]}; both quantities vanish for N\'{e}el SkXs \changed{by symmetry}.  We find that this factor is \changec{quantified} by the classical toroidal moment~\cite{hayami2014toroidal}
\begin{align}
  \vec{t} & = \frac{g \mu_{\mathrm{B}}}{2} \sum_{i} \vec{r}_{i} \times \vec{s_{i}}
  \propto \sin(\gamma) \,\vec{e}_z\label{eq:toroid_classic}
\end{align}
($\vec{r}_{i}$ position of spin $\vec{s}_{i}$ with respect to the skyrmion center). $\vec{t}$ is a pure real-space quantity given by the skyrmion helicity $\gamma$ (blue line in Fig.~\ref{fig:hel}). This easily accessible quantity successfully reproduces the functional dependence of $\alpha_{xy}$ (and also of $T_z$) on the helicity but fails to reproduce the proportionality factor because it does not depend on $E_{\mathrm{F}}$. Being a classical quantity, $\vec{t}$ can not explain the shape of $\alpha_{xy}(E_{\mathrm{F}})$ and $T_{z}(E_{\mathrm{F}})$. 

\changec{For a Bloch skyrmion ($\gamma=\pi/2$) the full $\alpha$ tensor is antisymmetric ($\alpha_{xy}=-\alpha_{yx}$) and has no longitudinal components. A N\'{e}el skyrmion ($\gamma=0$) exhibits only a longitudinal effect $\alpha_{xx}=\alpha_{yy}$ identical to $\alpha_{xy}$ of the Bloch skyrmion, since all spins are rotated about $\pi/2$ around the $\vec{z}$-axis}~\footnote{\changec{Similar to superexchange-driven magnetoelectricity in magnetic vortices~\cite{delaney2009superexchange}.}}. For antiskyrmions Eq.~\eqref{eq:toroid_classic} gives always zero. \changec{This is why the $\alpha$ tensor is symmetric and $T_z$ is zero in this case. Rotation of the sample always allows to diagonalize the tensor for antiskyrmion crystals since} $\gamma$ merely orients the two principle axes of an antiskyrmion\changec{, for which the texture points into opposite directions giving opposite longitudinal effects $\alpha_{xx}=-\alpha_{yy}$.}

\changec{The full tensor of the texture-induced magnetoelectric polarizability for a structural square lattice reads
\begin{align*}
\alpha(E_\mathrm{F})=\alpha_{xy}^{\mathrm{Bloch}}(E_\mathrm{F})\begin{pmatrix}
\cos(\gamma) & \sin(\gamma)  \\
-N_\mathrm{Sk}\sin(\gamma) & N_\mathrm{Sk}\cos(\gamma) 
\end{pmatrix}.
\end{align*}
The measurement of all tensor elements allows to determine topological charge $N_\mathrm{Sk}$ \emph{and} helicity $\gamma$ of an unknown skyrmion.}

\paragraph{Conclusion.} In this \changed{Rapid Communication}, we  established a complete scheme for the characterization of skyrmion crystals' topological charge and helicity (Fig.~\ref{fig:skyrmion}). Our findings on the topological Hall effect and the magnetoelectric effect are explained by quite simple pictures: a quantum Hall system and the classical toroidal moment of a spin texture, respectively.

Our prediction of the helicity-dependent magnetoelectric effect allows to discriminate N\'{e}el and Bloch skyrmions, without reverting to real-space imaging of their magnetic texture (which is in particular difficult for skyrmions arising at interfaces). For an electric field of $10^8\,\mathrm{V}/\mathrm{m}$ an additional in-plane magnetic moment of one-hundredth of $g\mu_B$ is induced per atom~\footnote{We used $\alpha_{xy}=0.1\, g \mu_{\mathrm{B}} e/at$ and typical values of $t=1\,\mathrm{eV}$ and $a=1\,\mathrm{nm}$. \change{The electrical field corresponds to the insulating case and has to be replaced by the corresponding electrical currents if the Fermi energy is located within a band.}}. The collapse of $\alpha_{xy}$ near van Hove singularities is a significant feature and could establish a new hallmark of the SkX phase: it is observable by shifting the Fermi energy (e.\,g., by a gate voltage or by chemical doping). 

\change{As shown in Fig.~\ref{fig:the}(f) and (g) as well as in the Supplemental Material~\cite{SupplementalMaterial}, the main claims of this \changed{Rapid Communication} depend qualitatively neither on skyrmion size, strength of the exchange interaction nor on the lattice geometry. The established scheme for discrimination (Fig.~\ref{fig:skyrmion}) is} \changeb{a general result, which is not limited to specific materials. All presented quantities arise solely due to coupling of `spin-full' electrons with the skyrmion texture and vanish in the absence of skyrmions. The skyrmion-induced contributions are distinguishable from the corresponding non-skyrmionic counter parts, e.\,g. the anomalous Hall effect in the presence of spin-orbit coupling, the `conventional' magnetization, and the `conventional' magnetoelectric effect in multiferroic materials~\cite{seki2012observation,seki2012magnetoelectric, kezsmarki2015neel,ruff2015multiferroicity}).}

\changeb{An experimental proof of the predicted magnetoelectric effect can simplest be done for a non-multiferroic material with a crystal symmetry that allows only for Bloch skyrmions (e.\,g. MnSi)~\cite{leonov2016properties}. The \changec{transverse} magnetoelectric effect arises purely due to toroidal order of the SkX and should be measurable in an isolated manner in such a material. The experiment can be conducted in analogy to that of Ref.~\onlinecite{saito2018evidence}, in which the \changed{metallic} coplanar toroidal magnet UNi$_4$B was investigated.}

\begin{acknowledgments}
\changed{We are grateful to Yukitoshi Motome, Gerrit E. W. Bauer and Annika Johansson for fruitful discussions.}
  This work is supported by Priority Program SPP 1666 and SFB 762 of Deutsche Forschungsgemeinschaft (DFG).
\end{acknowledgments}

\bibliography{short,MyLibrary}
\bibliographystyle{apsrev}

\end{document}